\documentclass[aps,prl,showpacs,amsmath,nofootinbib,amssymb,twocolumn,superscriptaddress,10pt,longbibliography]{revtex4-1}

\usepackage{array}
\usepackage{color}
\usepackage{multirow}
\usepackage{dcolumn}
\usepackage{bm}
\usepackage{epsf}
\usepackage{graphicx}
\usepackage{amsmath} 
\usepackage{dsfont} 
\usepackage{tikz}

\usepackage{hyperref}

\newcommand{\beq}{\begin{equation}}
\newcommand{\eeq}{\end{equation}}
\newcommand{\bea}{\begin{eqnarray}}
\newcommand{\eea}{\end{eqnarray}}

\begin{document}
\title{Constraints on neutrino-Majoron couplings using SN1987A data}

\author{Pilar Iváñez-Ballesteros}
\email{ivanez@apc.in2p3.fr}
\affiliation{Université Paris Cité, Astroparticule et Cosmologie, F-75013 Paris, France}

\author{M. Cristina Volpe}
\email{Corresponding author: volpe@apc.in2p3.fr}
\affiliation{CNRS, Université Paris Cité, Astroparticule et Cosmologie, F-75013 Paris, France}

\begin{abstract}
Neutrino decay to a lighter neutrino and a massless or almost massless (pseudo)scalar Goldstone boson remains of wide interest, as in the search for ultralight dark matter or of neutrinoless double beta-decay, and for its implications in astrophysics and cosmology.
Neutrino interactions with Majorons can affect the dynamics of supernovae and impact the emitted neutrino flux.
Using a three-neutrino framework and detailed supernova simulations, 
we perform the first likelihood analysis of the 24 neutrino events from SN1987A, including nonradiative decay in matter to a massless (pseudo)scalar boson like a Majoron. Focusing on the induced spectral distortions, we present bounds on the neutrino-Majoron couplings, as a function of the lightest neutrino mass, that are either complementary or competitive with current ones.
\end{abstract}
\date{\today}


\maketitle

\textit{Introduction.---}
Weakly coupled to matter, neutrinos are elusive particles that play an important role in astrophysics and cosmology.
They can influence the formation of large-scale structures and impact, e.g., the dynamics of massive stars, primordial and stellar nucleosynthesis in supernovae (SNe) and kilonovae, or the cooling of red giants and neutron stars. Neutrinos are also related to unresolved crucial issues, like the nature of dark matter, or the possibility of total lepton number violation if neutrinos are Majorana particles.

Bahcall, Cabibbo, and Yahil \cite{Bahcall:1972my} first suggested massive neutrinos could decay as an explanation for the $^{37}$Cl deficit observed in the pioneering experiment on solar neutrinos by Davis, Harmer, and Hoffmann \cite{Davis:1968cp}.
About a decade later, Chikashige, Mohapatra, and Peccei pointed out that 
the lepton number could be spontaneously broken globally, with an associated massless Goldstone boson, the {\it Majoron} \cite{Chikashige:1980qk}.
Soon after, triplets and doublets Majoron models were suggested \cite{Gelmini:1980re,Georgi:1981pg}, although later excluded by Z-boson decay \cite{Berezhiani:1992cd}. 

Astrophysical sources and the early Universe offer an exceptional possibility to search for non-standard neutrino properties and new physics.
In particular, core-collapse supernovae, releasing most of their gravitational energy as neutrinos, and the diffuse supernova neutrino background, formed by past supernova explosions, have a unique sensitivity to neutrino nonradiative two-body decay \cite{Fogli:2004gy,DeGouvea:2020ang,Ivanez-Ballesteros:2022szu}. In such a process, a new invisible particle, such as a Majoron, could be produced. For example, the lower limit on the lifetime-to-mass ratio $\tau/m > 3 \times 10^5$ s/eV (90 $\%$ C.L.) was obtained, using SN1987A events, for inverted neutrino mass ordering, improving previous constraints \cite{Ivanez-Ballesteros:2023lqa}.

Numerous works have provided bounds on the neutrino couplings to massive and massless Majorons. 
Bounds on the neutrino-Majoron coupling {$g_{ee}$ have been obtained from neutrinoless double beta decay ($0\nu\beta\beta$) experiments (see for example \cite{Brune:2018sab}).
Currently, the most stringent limits are provided by EXO-200 and NEMO-3, which obtained $g_{ee} < (0.4-0.9)\times 10^{-5}$ \cite{Kharusi:2021jez} and $g_{ee} < (1.6-3.0)\times 10^{-5}$ \cite{NEMO-3:2015jgm}, respectively.
Note that neutrino signals from Majoron dark matter could also be observed by dark matter detectors \cite{Garcia-Cely:2017oco}.
In the cosmological context, limits were obtained from big-bang nucleosynthesis \cite{Chang:2024mvg,Huang:2017egl} and the cosmic microwave background \cite{Escudero:2019gvw,Archidiacono:2013dua}. 
In astrophysics, effects of Majorons have been explored on the core-collapse supernova dynamics \cite{Dicus:1982dk, Kolb:1981mc, Fuller:1988ega} and cooling \cite{Farzan:2002wx}, as well as in the context of a future supernova \cite{Heurtier:2016otg} and the diffuse supernova neutrino background \cite{Akita:2022etk}. 

As for SN1987A, using the 24 neutrino events, constraints on neutrino couplings to Majorons were deduced by several studies
\cite{Kolb:1987qy,Kachelriess:2000qc,Farzan:2002wx,Tomas:2001dh,Choi:1987sd,Choi:1989hi, Heurtier:2016otg, Berezhiani:1989za,Brune:2018sab}. 
The majority exploited either the supernova cooling, the deleptonization arguments, or both. 
Indeed, according to the cooling argument, if Majorons were produced, they would modify the expected neutrino luminosity of about $3 \times 10^{53}$ ergs, released over 
the $\sim$10-second neutrino signal.
For the deleptonization, it is commonly assumed that the minimum lepton number of $Y_e \ge 0.375$ ($Y_e$ being the electron fraction) at the time of core bounce would ensure a successful supernova explosion. However, this value is based on the prompt supernova explosion mechanism \cite{Baron:1987zz}, ruled out in particular by the Bayesian analysis of Ref.~\cite{Loredo:2001rx}, later confirmed by \cite{Pagliaroli:2008ur}. On the other hand, bounds on neutrino-Majoron couplings were also derived from neutrino spectral distortions due to neutrino decay into massless Majorons \cite{Kachelriess:2000qc}.
Finally, bounds on massive bosons decaying to neutrinos were also extracted from the absence of high-energy neutrinos from SN1987A \cite{Fiorillo:2022cdq}.

In this letter, we perform a detailed $3\nu$ analysis of two-body neutrino nonradiative decay in matter, in relation to SN1987A. 
We investigate the spectral distortions induced by neutrino decay using modern one-dimensional supernova simulations, a detailed treatment of the neutrino decoupling region, and the temporal evolution of the neutrino emission for the first time.
With these inputs, we perform the first two-dimensional likelihood analysis of the 24 neutrino events from SN1987A including neutrino nonradiative decay in matter. We derive novel bounds on neutrino-Majoron couplings as a function of the lightest neutrino mass, for both neutrino mass orderings. Our 
constraints are complementary to the ones from neutrinoless double-beta decay and supernova cooling arguments, and competitive with those obtained from analyses using particle decays.

\textit{Neutrino decay to Majorons in supernovae.---}
The Lagrangian that describes the interactions of massless Majorons with neutrinos can be written as
\begin{equation}\label{eq:lagrangian}
    \mathcal{L}_{\rm int} \propto \sum_{i,j} g_{ij}\bar\nu_i\gamma_5 \nu_j J  ~ ,
\end{equation}
where $\nu_{i,j}$ with $i,j \in [1,3]$ and $J$ are the neutrino and the Majoron fields, respectively, and $g_{ij}$ the neutrino-Majoron couplings.
In Majoron models \cite{Schechter:1981cv}, the coupling matrix is proportional to the neutrino mass matrix that is diagonal, i.e., 
$g_{ij} \propto m_i \delta_{ij}$, in the mass basis. As a consequence, neutrino decay to Majorons in vacuum is suppressed.

For 3$\nu$ flavors, the proportionality $g_{ij} \propto m_i \delta_{ij}$ implies in particular that, using neutrino mass-squared differences measured in oscillation experiments, one can relate the $g_{ii}$ to the other two diagonal couplings in vacuum (here $i = 1$ or $i=3$ in the normal and inverted neutrino mass ordering respectively).
However, the effective masses that neutrinos acquire from their interactions with the medium~\cite{Berezhiani:1987gf} can induce non-zero off-diagonal terms $g_{ij}$ ($i \neq j$), allowing new interaction channels to open up and enhance neutrino decay in matter.

\begin{figure}
\begin{center}
\includegraphics[scale=0.4]{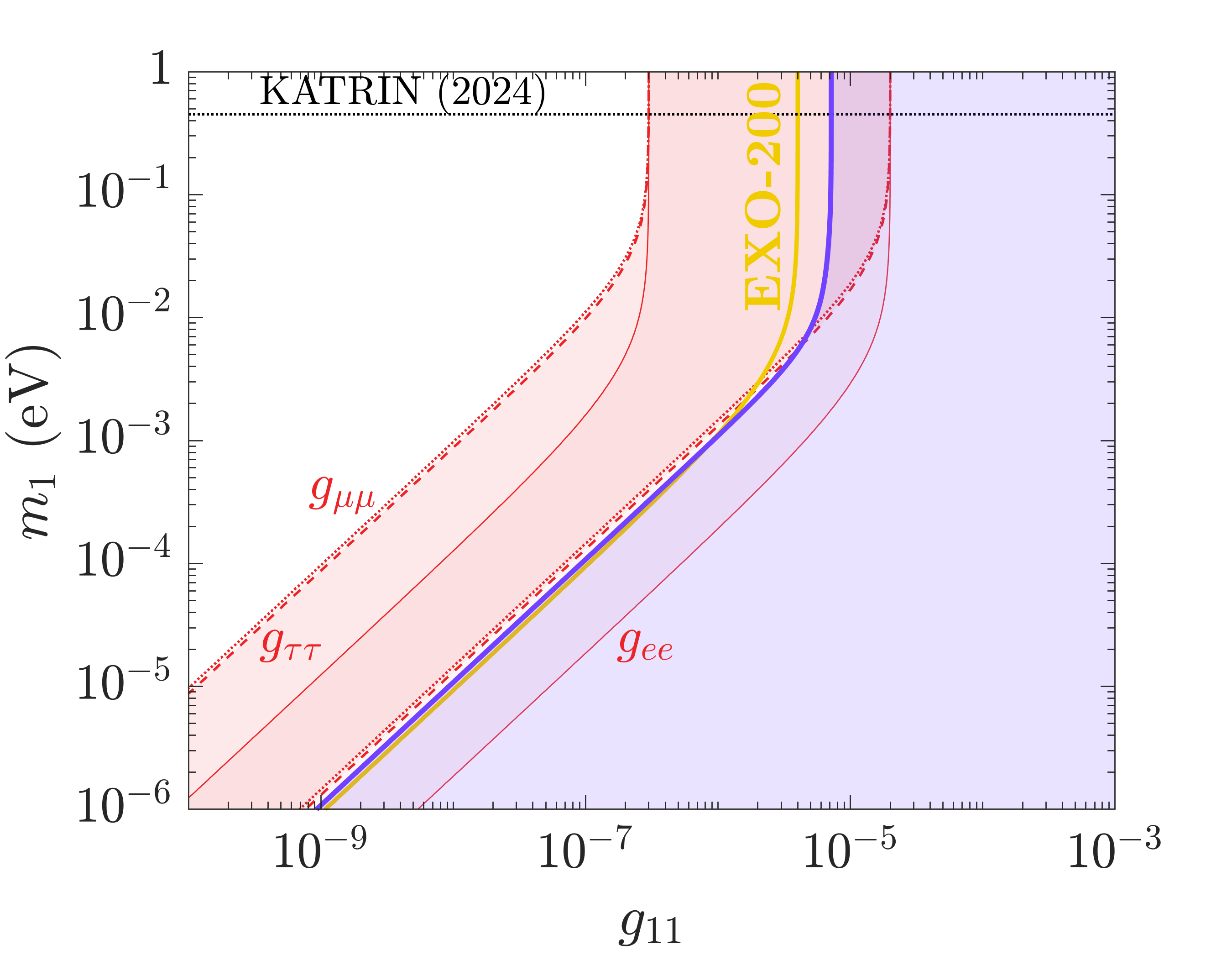}
\caption{Constraints on the neutrino-Majoron coupling in the $g_{11}-m_1$ plane from our SN1987A analysis considering normal ordering at 68\% CL (blue), compared with constraints from the Majoron luminosity argument (red) as per Ref.~\cite{Kachelriess:2000qc} and the most stringent limits from $0\nu\beta\beta$ decay (yellow solid line) \cite{Kharusi:2021jez}. The different red-shaded areas are obtained by requiring the limit from Ref.~\cite{Kachelriess:2000qc} on $g_{ee}$ (red solid line), $g_{\mu\mu}$ (red dotted line), and $g_{\tau\tau}$ (red dashed line). The current limit from the KATRIN experiment on the absolute neutrino mass is also shown \cite{Katrin:2024tvg}.}
\label{fig:g11-m1}
\end{center}
\end{figure}

In the standard framework of neutrino propagation in matter, the neutrino Hamiltonian includes the vacuum term, responsible for the established vacuum oscillations, and
the matter contribution. Indeed, neutrinos interact with the particles in the medium (mostly electrons, protons, and neutrons) through charged (CC) and neutral current (NC) interactions. While electron neutrinos ($\nu_e$) and antineutrinos ($\bar\nu_e$) experience both CC and NC interactions, non-electron neutrinos ($\nu_x$) antineutrinos ($\bar\nu_x$, $x = \mu, \tau$) are only affected by NC interactions. These can be accounted for through effective potentials (in the mean-field approximation). 
For neutrinos, they read $V_{\rm CC} = \sqrt{2} G_F n_B (Y_e + Y_{\nu_e})$ and $V_{\rm NC} = \sqrt{2} G_F n_B (-\frac{1}{2}Y_N + Y_{\nu_e})$, with $G_F$ the Fermi constant, $n_B$ the baryon density, and $Y_i = (n_i - \bar n_i)/n_B$ the particle number fraction with $i = e$ (electrons), $\nu_e$, and $N$ (neutrons). For antineutrinos, such potentials have opposite signs. 
For a given flavor $\alpha$, we will indicate $ V_{\alpha} = V_{\rm CC} + V_{\rm NC} $ or  $V_{\alpha} = V_{\rm NC} $ for the $\nu_e$ or $\nu_x$ respectively.

In the presence of matter, a convenient basis is the matter basis $|\Tilde\nu_i\rangle$, which instantaneously diagonalizes the full neutrino Hamiltonian and is related to the flavor basis through $| \nu_\alpha^{(\pm)} \rangle = \Tilde U^{(\pm)}_{\alpha i}  | \Tilde{\nu}_i^{(\pm)} \rangle$, with $\Tilde U^{(\pm)}$ the mixing matrix in matter. Here the superscripts $(\pm)$ denote helicity states, with positive (negative) helicity for antineutrinos (neutrinos). Close to the supernova core, the density is so high that the matter eigenstates can be identified with the flavor eigenstates (up to an arbitrary rotation in the $\nu_\mu$-$\nu_\tau$ subspace) \cite{Dighe:1999bi}. 
 
We now consider that neutrino decay to Majorons takes place when neutrinos escape the trapping region. Neutrino decay only produces spectral distortion once neutrinos fall out of thermal equilibrium.
Under the ultrarelativistic approximation, only helicity-flipping decays are allowed Eq.~\eqref{eq:lagrangian}. Since the matter can be identified with the rotated flavor basis, 
the total decay rate of the process $\nu_\alpha^{(\pm)} \to \nu_\beta^{(\mp)} + J$ reads \cite{Berezhiani:1987gf, Kachelriess:2000qc} 
\begin{equation} \label{eq:decayrate}
    \Gamma_{\alpha\beta} = \frac{g_{\alpha\beta}^2}{16\pi}\Delta V_{\alpha\beta} \ .
\end{equation}
One can see that the condition $\Delta V_{\alpha\beta} = V_\alpha - V_\beta > 0$, where the potential  $V_{\alpha, \beta}$ is the mean-field potential felt by a given flavor, in matter,  is necessary for the decay to occur. 
Note that the coupling matrix in the matter basis is related to the one in flavor basis by $\Tilde{g}_{ij} = U_{23}(\theta_{23}) g_{\alpha\beta} U_{23}(-\theta_{23}) = g_{\alpha^\prime\beta^\prime}$ where $\theta_{23}$ is one of the neutrino mixing angles \cite{Kachelriess:2000qc, Tomas:2001dh}.

In order to calculate the neutrino flux including neutrino decay, we need to consider in detail the energy and flavor-dependent regions where neutrinos evolve from trapped to free-streaming. 
On the one hand, beta processes ($\nu_e + n \leftrightarrow e^- + p$, $\bar{\nu}_e + p \leftrightarrow e^+ + n$) keep $\nu_e$ and $\bar{\nu}_e$ in thermal equilibrium with the medium up to a certain radius: the ``neutrinosphere". On the other hand, $\nu_x$ and $\bar{\nu}_x$ maintain thermal equilibrium with the medium mostly through nucleon-nucleon bremsstrahlung ($NN \leftrightarrow NN \nu\bar\nu$). This reaction freezes out at the ``energysphere". Yet $\nu_x$ and $\bar{\nu}_x$ remain trapped by NC collisions on nucleons ($\nu N \leftrightarrow \nu N$) up to the ``transportsphere"~\cite{Raffelt:2001kv}. 
Note that we do not consider the processes of Majoron decay and neutrino-Majoron scattering.
so that the bounds we derive are valid in the limit that Majorons are free-streaming \cite{Kachelriess:2000qc,Brune:2018sab,Heurtier:2016otg}. 
It is noted that the presence of inelastic neutrino scattering mediated by Majorons could modify the supernova dynamics, as investigated in the early work \cite{Fuller:1988ega}.
However, standard weak processes should dominate in the supernova dynamics, for the range of neutrino-Majoron couplings where the bounds we will present lie, for inverted neutrino mass ordering, while for normal mass ordering when $m_1 > 3 \times 10^{-3}$ eV. 
For the latter and very small $m_1$ values, further investigations beyond the free-streaming regime might be necessary (see eq. 6b of ref. \cite{Fuller:1988ega}).
Note that ref.\cite{Suliga:2024nng} also showed that the inclusion of neutrino-Majoron couplings could impact the supernova dynamics in the case of massive Majorons and for large values of the couplings, which is not what is considered in the present work.

Thus, when considering neutrino decay into Majorons, or a massless (pseudo)scalar boson in the inner core regions, 
for $\nu_{\alpha}$ or $\bar{\nu}_{\alpha}$ emitted from the energysphere, the survival probability is given by \cite{Kachelriess:2000qc}
\begin{multline} \label{eq:survival}
    N_{\nu_\alpha} (E) = \exp\Biggl\{ 
    - \int_{R_{{\rm ES}, \nu_\alpha}}^{R_{{\rm TS}, \nu_\alpha}} 
    \frac{dr^\prime}{v(E,r^\prime)} \Gamma_{\nu_\alpha}(r^\prime) \\
    - \int_{R_{{\rm TS}, \nu_\alpha}}^{\infty} 
    dr^\prime~\Gamma_{\nu_\alpha}(r^\prime)
    \Biggr\} \ , 
\end{multline}
where $E$ is the neutrino energy and $r'$ is the radial distance. The term $\Gamma_{\nu_\alpha} = \sum_{\beta = e, \mu, \tau} \Gamma_{\alpha\beta}$ represents the total decay rate of $\nu_\alpha$, and $R_{{\rm ES}, \nu_\alpha}$ and $R_{{\rm TS}, \nu_\alpha}$ are the radii of the energysphere and transportsphere, respectively.  

The first exponential term in Eq.~\eqref{eq:survival} accounts for the decay of neutrinos from their energysphere $R_{{\rm ES}, \nu_\alpha}$ to their transportsphere $R_{{\rm TS}, \nu_\alpha}$. In this region, neutrinos are still trapped and diffuse outward with an effective velocity $v(E, r) = \lambda(E, r)/( R_{{\rm TS}, \nu_\alpha} - R_{{\rm ES}, \nu_\alpha})$. The quantity $\lambda(E, r) = (\sigma(E)\rho(r))^{-1} $ is the mean free path which depends on the radius $r$, energy $E$, the medium density $\rho$ (that is also a function of time $t$) and the microscopic processes involved ($\sigma(E)$ is the corresponding cross-section). 
The second energy-independent contribution in the exponential implements the effect of decay after neutrinos start free-streaming, beyond the transportsphere. 

The two contributions in the survival probability Eq.~\eqref{eq:survival} differ depending on the 
neutrino species. Electron (anti)neutrinos fall out of thermal equilibrium and start free streaming at the neutrinosphere. 
Thus, Eq.~\eqref{eq:survival} can be simplified to only one integral inside the exponential, from the neutrinosphere to infinity.
On the contrary for muon and tau (anti)neutrinos, the energy-dependent contribution from the first term turns out to be important, as first pointed out by \cite{Kachelriess:2000qc}, and
needs careful treatment. We now turn to our calculations of the neutrino fluxes including neutrino decay and describe our inputs and computation of the two contributions in Eq.~\eqref{eq:survival}.

\textit{Fluxes of SN1987A neutrinos in presence of nonradiative decay in matter.---} 
In order to determine the spectra of SN1987A neutrinos on Earth, we need to calculate first the impact of neutrino nonradiative two-body decay and
then the one of the Mikheev-Smirnov-Wolfenstein (MSW) effect \cite{Wolfenstein:1977ue,Mikheev:1986wj}. Note that we do not consider here the possibility that sizable $\nu\nu$ NC interactions
producing (slow or fast) flavor conversion modes close to the supernova core, or flavor modification from shock waves and turbulence (see \cite{Volpe:2023met} for a review).

To determine the neutrino decay rates Eq.~\eqref{eq:decayrate} and survival probabilities Eq.~\eqref{eq:survival}, we need to model the neutrino fluences and
calculate how they evolve from the free-streaming regions to the stellar surface. 
To this aim, we used results on $\rho(t,r)$ and $Y_e(t,r)$ profiles from modern one-dimensional supernova simulations obtained by the Garching group~\cite{Garching, Fiorillo:2023frv}, for two different final neutron-star masses, $1.44~M_\odot$ and $1.62~M_\odot$, each with four different equations of state. For the neutrino flux parameters (luminosity, average energy and pinching parameter, we employ results from such simulations. Note that the fluences from these simulations are in good agreement with SN1987A data \cite{Fiorillo:2023frv}. 

As for the decay rates, for the profiles used and the relevant radii, the term $\Delta V_{\alpha\beta}$ from Eq.~\eqref{eq:decayrate} is positive only for the decay of antineutrinos to neutrinos (as was the case, e.g., in \cite{Kachelriess:2000qc}). Therefore, we only considered the decay of antineutrinos, and neutrinos were treated as stable. The antineutrino fluxes after decay become $\phi^d_{\bar\nu_\alpha} = N_{\bar\nu_\alpha} \phi^0_{\bar\nu_\alpha}$, where $\phi^0$ indicates the flux at the neutrinosphere for $\bar\nu_e$ and energysphere for $\bar\nu_\mu$ and $\bar\nu_\tau$. We used the neutrino fluxes obtained by the Garching group \cite{Garching, Fiorillo:2023frv}.

Concerning the survival probabilities, contrarily to \cite{Kachelriess:2000qc}, we performed a detailed analysis of the two contributions in Eq.~\eqref{eq:survival}}. We
considered time-dependent neutrino fluxes, matter density, and electron fraction profiles and computed the mean free paths for the different neutrino species, associated with the above-mentioned microscopic processes. For the calculations of the energy and flavor-dependent $R_{{\rm ES}, \nu_\alpha}$ and $R_{{\rm TS}, \nu_\alpha}$ we followed Ref.~\cite{Raffelt:2001kv}. As mentioned before, these spheres are time and energy dependent. However, we checked that for late times ($t \gtrsim 0.5$~s) and energies $\gtrsim 5$~MeV this energy dependence is not very strong. Therefore, we used average values of $R_{{\rm ES}, \nu_\alpha}$ and $R_{{\rm TS}, \nu_\alpha}$ over the neutrino energy range $E \in [7.5, 50]$~MeV relevant for our calculations.
 As for the second term in Eq.~\eqref{eq:survival} that we implement, this energy-dependent contribution is typically subdominant, as we have verified numerically.

For computational efficiency, we divided the supernova evolution into six time intervals and, for each interval, selected a representative time point (time templates). We calculated the survival probabilities at these times and applied them to the fluxes in the corresponding time ranges. In each interval, the survival probabilities were assumed to be constant. 
We checked that our results agreed with the work \cite{Kachelriess:2000qc} when considering only one time interval and the SN profiles from the old Wilson's simulations in \cite{Nunokawa:1997ct}. Then, we verified that our results converged when increasing the number of time templates from six to ten.
Afterward, we time-integrated the fluxes to obtain the fluences on Earth (the time information of the SN1987A events is not used in our likelihood analysis). More details on the calculations will be available in a forthcoming publication \cite{Ivanez-Ballesteros:2025ojj}.

After neutrinos start free-streaming and decay, they encounter the MSW resonances in the supernova envelope. These resonances occur at densities significantly lower than those required for the enhancement of neutrino decay to Majorons. Therefore, the MSW effect can be considered independently from the neutrino decay processes taking place in the dense regions. The antineutrino fluxes at the surface of the supernova become $\phi_{\bar\nu_1} = \phi^d_{\bar\nu_e} (\phi^d_{\bar\nu_{\tau^\prime}})$, $\phi_{\bar\nu_2} = \phi^d_{\bar\nu_{\mu^\prime}}$, $\phi_{\bar\nu_1} = \phi^d_{\bar\nu_{\tau^\prime}} (\phi^d_{\bar\nu_e})$ for normal (inverted) ordering. In these expressions, $\bar\nu_{\mu^\prime}$ and $\bar\nu_{\tau^\prime}$ represent the rotated flavor basis.

\begin{figure}
\begin{center}
\includegraphics[scale=0.4]{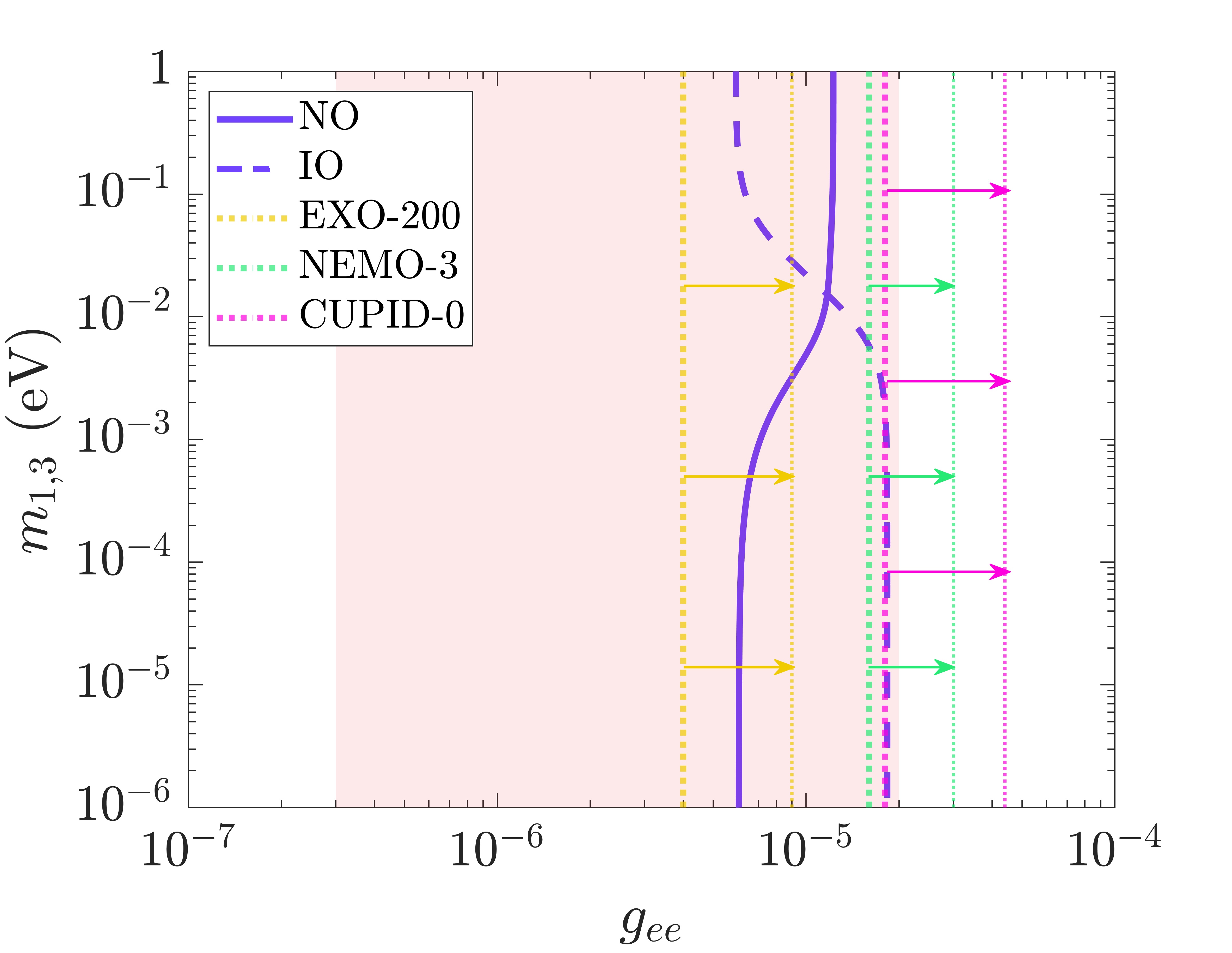}
\caption{Constraints at 90\% CL on the $g_{ee}$ neutrino-Majoron coupling as a function of the lightest neutrino mass obtained from our $3 \nu$ likelihood analysis for SN1987A events in normal (blue solid line) or inverted (blue dashed line) ordering. The red-shaded area indicates constraints from the Majoron luminosity argument~\cite{Kachelriess:2000qc}. 
Note that this limit assumes only one flavor at a time.
Additionally, we show limits from $0\nu\beta\beta$ decay (from more stringent to less): EXO-200~\cite{Kharusi:2021jez} (yellow dotted line), NEMO-3~\cite{NEMO-3:2015jgm} (green dotted line), and CUPID-$0$~\cite{CUPID-0:2022yws} (pink dotted line). For each bound, the two dashed vertical lines connected by an arrow show how each bound varies when implementing the uncertainty from the nuclear matrix elements.}
\label{fig:gee-m1}
\end{center}
\end{figure}

\begin{figure}
\begin{center}
\includegraphics[scale=0.4]{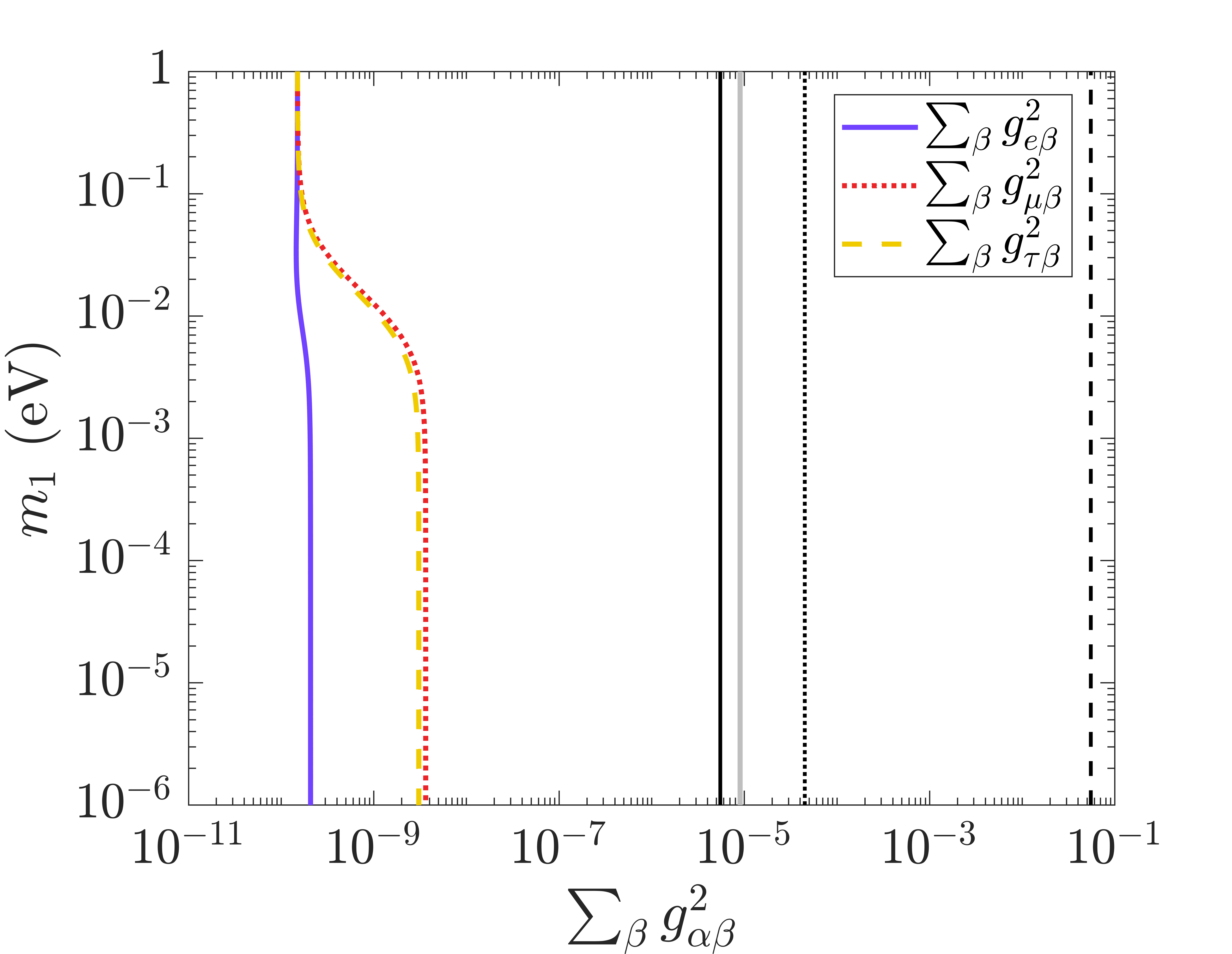}
\includegraphics[scale=0.4]{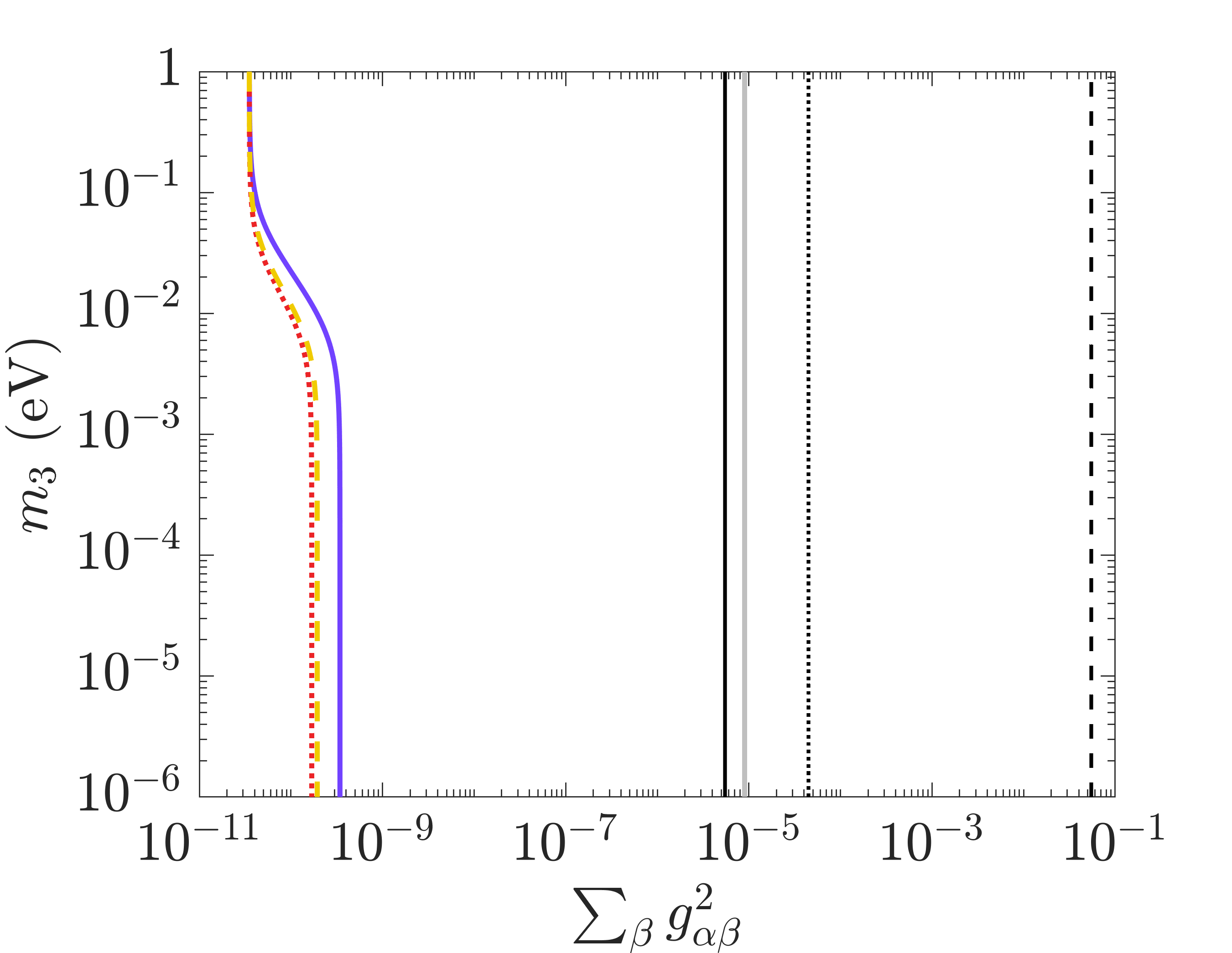}
\caption{Constraints at 90\% CL on $\sum_\beta g_{\alpha\beta}^2$ for $\alpha = e$ (blue solid line), $\mu$ (red dotted line), and $\tau$ (yellow dashed line) obtained from our SN1987A analysis in normal (upper) and inverted ordering (lower figure). As a reference, limits obtained from pion decay \cite{PIENU:2021clt} (gray solid line), kaon decay (black solid line), $\mu$ decay (black dotted line), and $\tau$ decay (black dashed line) \cite{Lessa:2007up}.}
\label{fig:gab-m1}
\end{center}
\end{figure}

\textit{Likelihood analysis of SN1987A data.---} 
Neutrinos from SN1987A, located 50~kpc away, were observed at three detectors: Kamiokande-II (2.14~kton), IMB (6.8~kton), and Baksan (0.28~kton)~\cite{Kamiokande-II:1987idp, Bionta:1987qt, Alekseev:1988gp}. These water Cherenkov detectors detected 11, 8, and 5 $\bar\nu_e$ events, respectively, via inverse beta decay (IBD). Further information about these events can be found in the Appendix of Ref.~\cite{Ivanez-Ballesteros:2023lqa}. At each detector, the $\bar{\nu}_e$ flux is given by $\phi_{\bar{\nu}_e} (E_{\nu}, L) = \sum_k \vert U_{ek}  \vert^2 \phi_{\bar{\nu}_k} (E_{\nu}, L) $ where $L$ is the SN1987A distance from Earth and $U_{ek} $ are matrix elements of the Pontecorvo-Maki-Nakagawa-Sakata matrix for which we use $\sin^2\theta_{12} = 0.307$, $\sin^2\theta_{23} = 0.547 (0.534)$, and $\sin^2\theta_{13} = 0.02$ in normal (inverted) ordering \cite{ParticleDataGroup:2022pth}. 
With these fluxes, we calculate the $\bar{\nu}_e$ events in each detector closely following Ref.~\cite{Ivanez-Ballesteros:2023lqa}, which provides a detailed description of the inputs and the detectors' responses.

We used a time-independent unbinned likelihood including statistical and systematic uncertainties following Refs. [10, 53] to extract information on the neutrino-Majoron couplings. For the likelihood analysis performed, we took as fit parameters $g_{11}~(g_{33})$ and $m_1~(m_3)$ in normal (inverted) ordering. The other two couplings in the mass basis were set as $g_{22} = g_{11}\sqrt{1+\Delta m^2_{21}/m_1^2}$ and $g_{33} = g_{11}\sqrt{1+\Delta m^2_{31}/m_1^2}$ for the normal ordering, with analogous expressions for $g_{11}$ and $g_{22}$ in the inverted ordering. 
We took for the squared-mass differences the following values, $\Delta m^2_{21} = 7.53\times10^{-5}$~eV$^2$ and $\Delta m^2_{32} = 2.437~(-2.519) \times10^{-3}$~eV$^2$ in normal (inverted) ordering \cite{ParticleDataGroup:2022pth}.

In order to find limits to the neutrino-Majoron coupling, we defined $\lambda(g_{11}, m_1) = \mathcal{L}(g_{11},m_1) / \mathcal{L}_{\rm max}$, where $\mathcal{L}(g_{11},m_1)$ is the likelihood function and $\mathcal{L}_{\rm max}$ is the maximum likelihood. The quantity $-2 \log[ \lambda(g_{11}, m_1) ]$ follows a $\chi^2$-distribution with two degrees-of-freedom (for sufficiently large samples)~\cite{Cowan:2010js}.

\textit{New constraints on the neutrino-Majoron couplings.---}
We performed our analysis considering eight different supernova simulations as mentioned above. Here we show the results of only one model, namely 1.44-SFHx. This model provides the best fit to SN1987A in the combined analysis performed in Ref.~\cite{Fiorillo:2023frv}. 
We checked that the late events that are not well captured by the modern SN1987A models used do not modify significantly our bounds. 

The results of our analysis on the $g_{11}-m_1$ plane (normal ordering) are shown in Fig.~\ref{fig:g11-m1}. The blue area indicates the region excluded by our analysis at 68\% confidence level (C.L.).  We studied how bounds vary when one uses 
the $1.44~M_\odot$ and $1.62~M_\odot$ models of Ref.44, with four different equations of state.
For $g_{11}$ that we show in Figure \ref{fig:gee-m1}, the dependence on the equation of state is small, while changing
from the $1.44~M_\odot$ and $1.62~M_\odot$  the bounds shift slightly on the right hand side. Note that the couplings $g_{22}$ and $g_{33}$
vary little when one changing models or equation of state. 

For comparison, we also show the region excluded by the Majoron luminosity argument: $3\times10^{-7} < |g_{\alpha\beta}| < 2\times10^{-5}$~\cite{Kachelriess:2000qc}. Note that the luminosity bounds are commonly derived assuming that only one flavor is present at a time.
The red-shaded areas are obtained by requiring this limit on $g_{ee}$ (red solid line), $g_{\mu\mu}$ (red dotted line), and $g_{\tau\tau}$ (red dashed line). Additionally, we translated the most stringent limit from $0\nu\beta\beta$ decay experiments, EXO-200~\cite{Kharusi:2021jez}, to the $g_{11} - m_1$ plane indicated by the yellow solid line. For the latter one can see that our exclusion curves on $g_{11} - m_1$ are compatible with the most stringent limits from $0\nu\beta\beta$ decay.

Figures~\ref{fig:gee-m1} and \ref{fig:gab-m1} show our results translated to limits in the flavor basis. Specifically, Fig.~\ref{fig:gee-m1} compares our limits for $g_{ee}$ in normal (blue solid line) and inverted ordering (blue dashed line) with those from $0\nu\beta\beta$ decay experiments (dotted lines) and the ones from Majoron luminosity (red-shaded area). The regions to the right of the lines are excluded. Our results improve on the limits set by NEMO-3 \cite{NEMO-3:2015jgm} and CUPID-0 \cite{CUPID-0:2022yws}. Only the limits set by EXO-200 \cite{Kharusi:2021jez} are tighter than the results obtained in this study. Note, however, that the EXO-200 bounds shown are the tightest ones when implementing the uncertainties from the nuclear matrix elements. Additionally, such limits become much looser when considering other Majoron decay models.

Finally, limits were also obtained on $\sum_\beta g_{\alpha\beta}^2$ for $\alpha, \beta = e$, $\mu$, and $\tau$ using meson and lepton decays. Fig.~\ref{fig:gab-m1} presents a comparison of these with our results for $\sum_\beta g_{e\beta}^2$ (blue solid line) that improve by nearly 5 orders of magnitude over those obtained from kaon decay \cite{Lessa:2007up} (black solid line) and pion decay \cite{PIENU:2021clt} (gray solid line). One can also see that the limits on $\sum_\beta g_{\mu\beta}^2$ are improved (red dotted line) by 4 orders of magnitude compared to those obtained from $\mu$ decay \cite{Lessa:2007up} (black dotted line). The less stringent limits on $\sum_\beta g_{\tau\beta}^2$ derived from $\tau$ decay \cite{Lessa:2007up} (black dashed line) are improved by approximately 7 orders of magnitude (yellow dashed line). 

\textit{Discussion and conclusions.---}
We have analyzed the 24 neutrino events from SN1987A including the possibility that neutrinos nonradiatively decay in matter, producing a (pseudo)scalar goldstone boson like a Majoron. 
Our work goes beyond previous analyses in several respects. First, we considered a 3$\nu$ framework instead of an effective two-flavor one, as done in previous works which set $g_{33}$ and $\theta_{13}$ equal to zero. Furthermore, our analysis uses profiles from state-of-the-art supernova simulations, which agree with SN1987A observations. We also considered the time dependence of these profiles, incorporated a time-dependent treatment of neutrino decay in matter, and determined the radii of the transportsphere and energysphere, aspects that were completely absent in previous works. 
Moreover, we performed, for the first time, a likelihood analysis of SN1987A data to obtain limits on the coupling between massless Majorons and neutrinos,
from the spectral distortions induced by the decay. 

As a result of these advancements, our bounds on $g_{11}$ (Fig.~\ref{fig:g11-m1}) are about one order of magnitude tighter than the limits quoted in Ref.~\cite{Kachelriess:2000qc} from the neutrino spectra, and we obtained improved limits on $g_{22}$ and new bounds on $g_{33}$. Moreover, our limits on the neutrino-Majoron couplings are competitive with limits obtained from laboratory experiments. Our results for $g_{ee}$ are surpassed only by the most stringent constraint from $0\nu\beta\beta$ decay obtained by EXO-200. However, $0\nu\beta\beta$ decay can only probe the $g_{ee}$ element of the coupling matrix. In contrast, our results also allow us to set limits on other elements of the coupling matrix. Furthermore, as shown in Fig.~\ref{fig:gab-m1}, our results on $\sum_\beta g_{\alpha\beta}^2$ improve by $4 - 7$ orders of magnitude the limits obtained from analyses of meson and lepton decays \cite{PIENU:2021clt,Lessa:2007up}.

In conclusion, the present investigation based on spectral distortions of the SN1987A neutrino events has allowed us to establish either tighter or novel limits on the neutrino-Majoron couplings. In the fiuture, using more advanced supernova simulations, such as three-dimensional models, will improve our description of supernova interiors making the results even more robust. Also, a fully consistent treatment of neutrino-Majoron scattering and Majoron decay would extend the current bounds to include the trapping regime, while it would require SN simulations including Majorons as well. 

This analysis provides a valuable framework for analyzing neutrino data from future supernovae which could yield even tighter constraints on the interactions of Majorons and neutrinos. 
While we have considered a massless boson as a neutrino decay product, this analysis is also valid for almost massless (pseudo)scalar bosons. Therefore the present work constitutes a step forward in establishing solid limits on potential ultralight dark matter candidates using neutrinos from core-collapse supernovae in our Universe.

The authors wish to thank the Galileo Galilei Institute for Theoretical Physics for the hospitality, the Masterproject NUFRONT, and I.N.F.N. for partial support during the completion of this work. We also thank H.T. Janka for providing useful information and access to the Garching archives.

\bibliography{references}

\end{document}